# Contention-based Grant-free Transmission with Independent Multi-pilot Scheme

Zhifeng Yuan, Weimin Li, Zhigang Li, Yuzhou Hu, Yihua Ma
ZTE Corporation, Shenzhen, China
Email: {yuan.zhifeng, li.weimin6, li.zhigang4, hu.yuzhou, yihua.ma}@zte.com.cn

*Abstract*—Contention-based grant-free transmission is very promising for future massive machine-type communication (mMTC). In contention-based transmission, the random pilot collision is a challenging problem. To solve this problem, multiple pilots scheme is used to reduce the pilot collision probabliltiy. However, the existing work on multiple pilots relies on the low correlation of spatial channels, limiting its applicability. In this paper, an independent multi-pilot scheme is proposed, which utilizes the diversity of multiple pilots and is not limited by the spatial correlation. The receiver employs interference cancellation for both data symbols and multiple pilots to ensure the performance. The simulation results also show that the proposed independent multi-pilot scheme can significantly improve the BLER performance and increase the number of simultaneous access users.

*Keywords—grant-free, independent multi-pilot, pilot collision, contention-based transmission*

## I. INTRODUCTION

In beyond 5G (B5G) wireless communication system, massive Machine-Type Communication (mMTC) will be one of the most important scenarios [1, 2]. The mMTC scenario features massive access devices, small data packets, low transmission rate and sporadic communication. In addition, the access devices should be characterized by low complexity, power-saving and low cost. The grant-free access scheme [3, 4], which allows user equipment (UE) to transmit data autonomously without the need to send scheduling request and wait for dynamic scheduling, has been proved to be a promising solution to satisfy the requirements of mMTC scenario with the following advantages: (1) saving signaling overhead; (2) reducing transmission latency: The complex random access and resource grant procedure are no longer needed; (3) reducing power consumption: The terminal devices can be in idle state for a long time, and can immediately switch to transmitting state when data arrives. At low transmission rate, non-orthogonal multiple access (NOMA), with the ability to exploit near-far advantage, has a higher capacity than orthogonal multiple access, and can support more access devices on the same transmission resources [5, 6]. The combination of grant-free and NOMA can solve the problems of connection density, signaling overhead, terminal complexity and power consumption, which is suitable to the B5G mMTC scenario.

In the mMTC scenario with massive connections and sporadic small data packets, reserving dedicated transmission resource for each connection is unrealistic, resulting in different UEs sharing the same wireless resource block. In autonomous grant-free transmission, each UE autonomously selects transmission resources including pilot sequence in a contention-based manner. Inevitably, multiple UEs may select the same pilot sequence known as pilot collision, which will severely degrade the system performance [7, 8]. For traditional transmission scheme where a single pilot sequence and a data payload are included, the pilot collision probability can be expressed as [9]

$$P = 1 - \frac{A_N^K}{N^K} \quad (1)$$

where $N$ is the size of the defined pilot pool containing orthogonal pilot sequences, $K$ is the number of concurrent UEs, and $A_N^K = N!/(N-K)!$ is the number of permutations of $N$ items taken $K$ at a time. It can be derived that, as the number of concurrent UEs $K$ increases, the pilot collision probability rises rapidly.

Increasing the number of pilot sequences to reduce the pilot collision probability is a common approach [10, 11]. However, as the number of orthogonal pilot sequences increases, the length of the pilot sequence also needs to be increased, which consumes more time-frequency resources. With limited time-frequency resources, the resources available for data are less, affecting the efficiency of data transmission. In addition, the complexity of blind detection for a longer pilot sequence during active user detection (AUD) also increases significantly. Another approach to solve the pilot collision problem is a data-only based grant-free transmission scheme [12, 13], which directly removes the pilot signal, resulting in no pilot collision and overhead. In this scheme, advanced blind multi-user detection (MUD) techniques without pilot-based channel estimation, such as blind receive beamforming without spatial channel information, blind activity detection based on the second order moment of the data symbols, bind MMSE de-spreading, and blind equalization via partition-matching method, are required.

In this paper, an independent multi-pilot (IMP) scheme is proposed, which can significantly reduce the pilot collision probability, with moderately increased blind detection efforts. Different from the traditional single pilot (TSP) scheme, the proposed scheme uses multiple pilot sequences which are selected or generated independently, thus uncorrelated, and collision occurring on all pilots is obviously lower than that in the TSP scheme under the same pilot resource overhead. Unlike the multiple pilots scheme relying heavily on the low correlation of the accessing users' spatial channels [14], which limits its applicability, the proposed IMP scheme works for a more general case without any requirement of spatial correlation. The simulation results show that both the Block Error Rate (BLER) performance and the number of access UEs are significantly improved at the cost of increased decoding attempts per user. The article is organized as follows. In Section II, the independent multi-pilot scheme including the multi-pilot based transmitter and receiver is elaborated in detail, then the pilot collision probability, channel estimation precision, receiver complexity, and the number of independent pilots are also discussed. Section III provides the performance evaluation results together with the comparison between the independent multi-pilot scheme and traditional single pilot scheme. The conclusion is presented in Section IV.

## II. THE INDEPENDENT MULTI-PILOT SCHEME

### A. Multi-pilot and Transmitter Design

Different from the traditional scheme configured with single pilot, the IMP scheme uses multiple independent pilots under the same pilot resource overhead. One typical resource sharing by the multiple pilots is shown in Fig. 1, where each pilot occupies a disjoint sub resource block of the pilot resource. In principle, the lengths or the sequence types of the pilots are not necessarily identical. However, multiple pilots of equal length and identical sequence type are beneficial to minimize the collision probability, channel estimation error and implementation complexity.

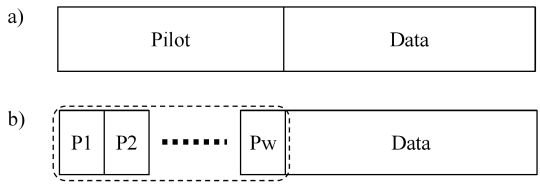

Fig. 1. a) Traditional single pilot scheme; b) Independent multi-pilot scheme

Taking $w$ independent pilots of equal length and identical sequence type as an example, each UE independently selects $w$ pilot sequences from a predetermined pilot pool and maps them to the transmission resources together with the data symbols in one-shot transmission. The total overhead of $w$ independent pilots is the same as that of the single pilot in traditional scheme. Assuming that the size of the pilot pool of the TSP scheme is $N_{TSP}$ (i.e., $N_{TSP}$ orthogonal pilot sequences of length $N_{TSP}$), and the $w$ independent pilots use disjoint resources, the size of the pilot pool of the IMP scheme $N_{IMP}$ is therefore $N_{TSP}/w$ (i.e., $N_{TSP}/w$ orthogonal pilot sequences of length $N_{TSP}/w$).

For IMP-based transmission, the probability of collision on all $w$ pilots $P_{IMP}$ would be significantly lower than that in TSP-based scheme $P_{TSP}$ provided $N_{TSP}$ is not too small, which is a reasonable assumption because $N_{TSP}$ in general needs to be large enough to ensure an acceptable collision rate for traditional scheme. Considering the case of two UEs randomly selecting pilot sequences from the predefined pilot pool, $P_{TSP}$ is $1/N_{TSP}$, $P_{IMP}$ is $(1/N_{IMP})^w = (w/N_{TSP})^w$. Assuming that two independent pilots are utilized, i.e. $w = 2$, $P_{IMP}/P_{TSP}$ is $4/N_{TSP}$, which means that IMP-based scheme has lower collision probability when $N_{TSP}$ is larger than 4, and the probability would be rapidly reduced with the increasing of $N_{TSP}$. Therefore, for IMP-based transmission, the detection of a given UE's data can depend on at least one of its pilots that does not collide with other UEs', and better performance can be expected due to the lower pilot collision probability.

To improve the detection performance of other UEs whose pilots are collided with the UEs that have been decoded, interference cancellation (IC) on pilot signals is required to be implemented. That means the receiver should know the pilot sequences the UE used. To realize this, putting the information of multiple pilot sequences into the data payload seems a straightforward solution, such that once a given UE's data is successfully decoded, all the pilot sequences used by the UE can be determined and then IC on the multiple pilots can be performed. However, this solution would incur overhead. Alternatively, some given bits in codeword, which in general are independent or uncorrelated of each other, can be used to determine the multiple pilot sequences. Concretely, if $N_{IMP}=2^m$, $w*m$ coded bits can be used to select the $w$ independent pilot sequences from the pilot pool. In this way, once the coded bits pass the CRC check, the indexes of $w$ pilot sequences can be determined.

### B. Receiver Design

The receiver of the IMP-based grant-free transmission scheme is mainly composed of two parts: blind MUD and codeword level IC. The receiver has no prior information on the number of access UEs and the pilots selected by these UEs. Therefore, the receiver has to perform blind detection including AUD and channel estimation, MMSE equalization, demodulation and decoding, etc., as shown in Fig. 2.

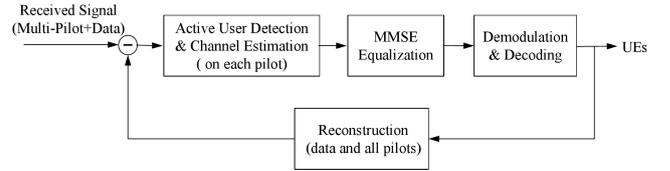

Fig. 2. Receiver processing diagram

For convenience, the subsequent description takes two independent pilots as an example. The receiver can perform blind MUD by using two independent pilots in parallel, and then perform IC on both the pilots and data symbols, as shown in Fig. 3a). It also can use one pilot to perform MUD and IC at first, then use another pilot to perform MUD and IC, i.e. two pilots are processed serially, as shown in Fig. 3b). If parallel processing procedure shown in Fig. 3a) is adopted, one UE may be successfully decoded based on both pilots. However this UE's signal can not be cancelled twice, the receiver therefore has to determine which channel estimation derived from P1 or P2 should be used to reconstruct the UE's received symbols. The receiver can also process the channel estimations on both pilots to obtain a weighted channel estimation for reconstruction. In another way, transmitted symbols of all decoded UEs can play the role of 'pilot' to derive refined channel estimations, which then can be used for more accurate received symbols reconstruction. This method can be called data-aided channel estimation, which is quite beneficial for contention-based grant-free transmission and will be discussed in detail in latter sections.

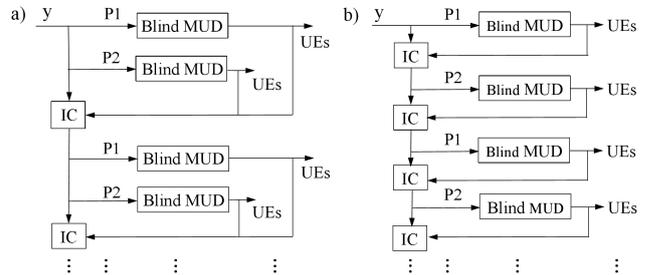

Fig. 3. a) Parallel detection based on two pilots; b) Serial detection based on two pilots

Assume $K$ active UEs are simultaneously transmitting data in contention-based grant-free manner, and each UE independently selects two pilot sequences from the defined orthogonal pilot resource pool of size $N_{IMP}$,

$\mathbf{Z} = \{\mathbf{z}_1, \mathbf{z}_2, ..., \mathbf{z}_{N_{IMP}}\}$, $\mathbf{z}_{N_{IMP}} \in C^{N_{IMP}}$. The received pilot symbols on the $i$-th pilot can be expressed as

$$\mathbf{y}_i = \sum_{k=1}^{K} h_{i,k} \mathbf{z}_{n,i,k} + \mathbf{n}_i, \quad i = 1, 2 \quad (2)$$

where $\mathbf{z}_{n,i,k}$ represents that UE $k$ randomly selects a pilot sequence $\mathbf{z}_n$ on the $i$-th pilot, $\mathbf{z}_n \in \mathbf{Z}$; $h_{i,k}$ is the channel coefficient of UE $k$ on the $i$-th pilot, here flat fading channel is assumed for simplicity; $\mathbf{n}_i$ is the additive white Gaussian noise (AWGN), $\mathbf{n}_i \sim CN(0, \sigma^2 I)$.

1) Blind detection

By a hypothesis testing on all possible pilot sequences, blind detection of active users' pilot sequences can be performed, which can be expressed as

$$\mathbf{z}_x^* \mathbf{y}_i = \sum_{k=1}^{K} h_{i,k} \mathbf{z}_x^* \mathbf{z}_{n,i,k} + \mathbf{z}_x^* \mathbf{n}_i, \quad x = 1, ..., N_{IMP} \quad (3)$$

where $(\cdot)^*$ represents conjugate transpose. By using the orthogonality of pilot sequences, i.e., $\mathbf{z}_x^* \mathbf{z}_n = 0, x \neq n$, the following can be derived for UE $k$

$$\mathbf{z}_n^* \mathbf{y}_i = h_{i,k} \mathbf{z}_n^* \mathbf{z}_n + \mathbf{z}_n^* \mathbf{n}_i \quad (4)$$

Then normalized detection result, which is also the channel estimation of UE $k$ on the $i$-th pilot, can be obtained by

$$\hat{h}_{i,k} = \frac{\mathbf{z}_n^* \mathbf{y}_i}{\mathbf{z}_n^* \mathbf{z}_n} = h_{i,k} + \frac{\mathbf{z}_n^* \mathbf{n}_i}{\mathbf{z}_n^* \mathbf{z}_n} \quad (5)$$

By setting appropriate detection threshold, active user on the $i$-th pilot can be identified based on the normalized detection results. The threshold should be set considering the influence of interference and noise, and to realize a trade-off between false alarm and miss detection. Assuming that $S$ users are identified, based on the channel estimations of these users, MMSE based equalization can be implemented to derive the detected data symbols of each user, which then can be sent to demodulator and decoder to obtain the data bits transmitted by corresponding users. CRC check can be used here to judge whether the data of a user is correctly decoded. As discussed before, information of multiple pilot sequences as well as identification number of a user can be carried in the data payload. So from the correctly decoded data bits, access user can be determined together with its multiple pilots, which would help the interference cancellation process.

2) Interference Cancellation

Pilot collision has a strong impact on the performance of contention-based grant-free transmission. If two or more UEs select the same pilot sequence, only one pilot sequence can be detected, and the channel estimation based on this pilot would be the sum of channels experienced by these UEs. If there are large disparity between the collided UEs' received power, the strongest UE may be successfully decoded. In order to detect the remaining UEs, the data symbols as well as the pilot signal of the successfully decoded UEs should be reconstructed, and cancelled from the received symbols, then blind MUD can be performed iteratively.

For IMP-based scheme, it should be noted that IC should be performed for all pilots. The two pilot sequences selected by a decoded UE can be determined according to the $2m$ bits in the data part. Assuming that $Q$ UEs are successfully decoded, and taking serial detection based on two pilots shown in Fig. 3b) as an example, the IC for pilots and data can be respectively expressed as

$$\mathbf{y}_p' = \mathbf{y}_p - \sum_{q=1}^{Q} \tilde{h}_{i,q} \mathbf{z}_{v,p,q}, \quad p = 1, 2 \quad (6)$$

$$\mathbf{y}_d' = \mathbf{y}_d - \sum_{q=1}^{Q} \tilde{h}_{i,q} \mathbf{d}_q \quad (7)$$

where $\mathbf{z}_{v,p,q}$ represents the pilot sequence $\mathbf{z}_v$ used by UE $q$ on $p$-th pilot, $\mathbf{z}_v \in \mathbf{Z}$; $\tilde{h}_{i,q}$ is the filtered channel estimation based on $\hat{h}_{i,q}$ of UE $q$ on the $i$-th pilot, it means that channel estimation on the $i$-th pilot would be used for cancellation on both pilots as well as data part; $\mathbf{d}_q$ is the reconstructed modulated data symbols of UE $q$.

Based on $\mathbf{y}_p'$ and $\mathbf{y}_d'$, the next round of blind MUD can be performed. The process can be iterated until no active UE can be identified or no new UE can be successfully decoded.

Generally, the accuracy of pilot-based channel estimation is limited by the pilot power, which is more obvious for IMP-based scheme because each pilot's power is reduced. It would cause large residual error and affect the detection of remaining UEs. In addition, as mentioned above, when pilot collision exists, the channel estimation is the sum of channels of muiltple UEs, IC based on that would be quite inaccurate, and the remaining weaker collided UEs may not be detected subsequently, resulting in miss detection.

To improve the accuracy of IC, data-aided channel estimation can be utilized, where all the transmitted symbols (including the pilot symbols and modulated data symbols) of decoded UEs can be employed as 'pilot' to derive refined channel estimations for these UEs through least square method [12]. It can be expressed as

$$\hat{\mathbf{h}} = (\mathbf{D}^* \mathbf{D})^{-1} \mathbf{D}^* \mathbf{y} \quad (8)$$

where $\mathbf{y}$ is a vector composed of the received symbols, $\mathbf{D} = [\mathbf{D}_1, \mathbf{D}_2, ..., \mathbf{D}_Q]$ is a matrix composed of the reconstructed transmitted symbols of all the $Q$ decoded UEs till this round, $\hat{\mathbf{h}} = [\hat{h}_1, \hat{h}_2, ..., \hat{h}_Q]^T$ is a vector composed of the channel estimations for these $Q$ UEs.

It can be seen from [13] that data-aided channel estimation would be more accurate with the increasing number of successfully decoded UEs, and it is more suitable for the case with pilot collision as the data of different UEs can be considered uncorrelated. Therefore, data-aided channel estimation in IC can help to minimize the residual error and reduce the miss detection rate, which are particularly important for the weaker UEs.

C. Pilot Collision Probability

More discussions are provided here to illustrate the advantages of IMP scheme in reducing pilot collision probability. The case with two UEs has been discussed before, while for the case with more UEs, the pilot collision analysis would be a little complex. Fig. 4 shows a case with

3 UEs ($K = 3$), wherein UE2 collides with others on both pilots, such that it cannot be decoded in the first round of blind MUD. However, UE3 could be successfully decoded based on its first pilot 'z4' which does not collide with others. Then the second pilot 'z5' of UE3, although collides with that of UE2, can be determined according to the $2m$ bits in the decoded data. Similarly, UE1 could be successfully decoded based on its second pilot 'z7'. After cancellation for UE1 and UE3, the receiver can perform next round of detection, where UE2 would be decoded. It can be seen that, although UE2 collides with others on both pilots, it finally can still be decoded. This case can be regarded as a solvable pilot collision case, and will not affect the BLER performance considerably.

|  | P1 | P2 |  |
|---|---|---|---|
| UE1 | z3 | z7 | Data $d1$ |
| UE2 | z3 | z5 | Data $d2$ |
| UE3 | z4 | z5 | Data $d3$ |

Fig. 4. One case of pilot collision with 3 access UEs

Two UEs colliding with each other on both pilots is the main case that affects the performance. For example, if UE1 shown in Fig. 4 selected 'z5' on pilot P2, only UE3 could be successfully decoded, but both UE1 and UE2 can not be decoded as they collide with each other on both pilots. The collision probability of this case for $K = 3$ can be expressed as

$$P = (C_K^2 A_N^2 A_N^2 + 2 C_K^2 A_N^2 A_N^1) / N^{2K} \quad (9)$$

where $N$ is the size of the pilot pool of the IMP scheme, i.e., $N = N_{\text{IMP}}$, and $C_N^K = \dfrac{N!}{K!(N-K)!}$ is the number of combinations of $N$ items taken $K$ at a time. Further, Eq. (9) can be approximated as

$$P \approx C_K^2 \frac{1}{N^2} \quad (10)$$

The probability in Eq. (10) can be regarded as the number of combinations for randomly selecting two UEs from three concurrent UEs multiplied by the probability of collision on both pilots for the case with two UEs. Generalizing Eq. (10) to $K > 3$ UEs seems reasonable.

Assuming that the size of pilot pool of the TSP scheme is $N_{\text{TSP}} = 24$, and the size of pilot pool of the IMP scheme configured with two pilots is $N_{\text{IMP}} = 12$, the pilot collision probability is about 12% for TSP scheme with 3 concurrent UEs according to Eq. (1), while according to Eq. (10) the pilot collision probability would reduce to 2%. The probability of three UEs colliding with each other on both pilots at the same time is negligible. For more UEs, the pilot collision probability can be analyzed similarly.

*D. Accuracy of Channel Estimation*

For IMP scheme, the total overhead and energy of $w$ pilots is the same as that of single pilot in the TSP scheme. Therefore, the energy of one pilot in the IMP scheme is only $1/w$ of that in the TSP scheme, which leads to lower channel estimation accuracy. Specifically, for the case with two pilots, the channel estimation accuracy will decrease by 3dB; and for the case with three pilots, the channel estimation accuracy will decrease by 4.77dB. To balance the channel estimation accuracy and demodulation performance, transmit power boost could be considered on the multiple pilots for IMP-based transmission.

*E. Receiver Complexity*

As the receiver needs to perform MUD based on each pilot for IMP scheme, the complexity would be increased. If the parallel detection procedure shown in Fig. 3a) is employed, the receiver may repeatedly decode the same UE on multiple pilots. While for the serial detection procedure shown in Fig. 3b), by using interference cancellation after detection on each pilot, decoding the same UE on multiple pilots can be avoided in some extent, which would save complexity. Some complexity comparison results are provided in Section III for reference.

*F. Number of Independent Pilots*

The number of independent pilots is closely related to the pilot collision probability, channel estimation accuracy and receiver complexity. As discussed above, increasing the number of pilots can reduce the pilot collision probability and improve the system performance in collision-limited scenarios. However, the more the number of pilots, the lower the accuracy of channel estimation. This will not only degrade the demodulation performance, but also lead to large residual error of interference cancellation, which directly affects the detection of the weaker UEs. The receiver complexity also increases with the number of pilots. Therefore, the determination of the number of pilots requires a comprehensive consideration of the system performance and complexity.

III. NUMERICAL RESULTS

The performance of the proposed IMP scheme is evaluated by link-level simulation, and the comparison with the TSP scheme is also provided. The simulation parameters are shown in Table 1.

TABLE I. SIMULATION ASSUMPTION

| Parameter | Value |
|---|---|
| Carrier frequency | 0.7 GHz |
| System bandwidth | 10 MHz |
| Allocated bandwidth | 6 PRBs |
| Modulation and Coding scheme | QPSK, LDPC |
| Size of transport block | 160 bits |
| Length of CRC | 16 bits |
| Antenna configuration | 1Tx, 2Rx |
| Channel model and delay spread | TDL-A 30ns |
| Receiver algorithm | MMSE-IC |

The allocated transmission resource is 6 consecutive physical resource blocks (PRBs) in the frequency domain, and 14 OFDM symbols in the time domain. There are 12 subcarriers per PRB and subcarrier spacing is 15 kHz. The Demodulation Reference Signal (DMRS) structure of TSP scheme can refer to DMRS type 2 in 5G new radio (NR). The difference is that the DMRS structure of TSP schneme is more sparser in the frequency domain for a larger number of orthogoal reference signals to increase the multiplexing capacity or support more access UEs. Multipe pilots of IMP scheme are mapped to independent, non-overlapping, and equal-sized sub-resource blocks in the frequency domain. Two kinds of pilot overhead are considered. In the first one,

2 OFDM symbols are for pilot, i.e. the pilot overhead is 1/7, thus 24 orthogonal reference signals can be configured on each PRB for TSP scheme, while for IMP scheme, 2 pilots with 12 orthogonal reference signals per pilot can be configured on each PRB, or 3 pilots with 8 orthogonal reference signals per pilot can be configured on each PRB. In the other one, 4 OFDM symbols are for pilot, i.e. the pilot overhead is 2/7, and the number of orthogonal reference signals doubles for each scheme. The channel model is TDL-A with delay spread of 30 ns. The long term receiving SNR of all access UEs are equal. In the receiver, the serial detection procedure shown in Fig. 3b) is used, and pilot based realistic channel estimation is employed in the detection and interference cancellation. To reduce the IC residual error and improve the detection performance, reconstructed data aided channel estimation is also considered for IC in the simulation.

### A. Pilot-based Channel Estimation Used in IC

In this subsection, pilot-based channel estimation is used in IC. The comparison of the BLER performance between the TSP scheme and the IMP scheme configured with 2 pilots under 1/7 pilot overhead is shown in Fig. 5. From the results, it can be seen that, the BLER performance of the IMP scheme is better than that of the TSP scheme in general. At BLER=0.1, the IMP scheme can support more than 8 simultaneous access UEs, which is about twice the number of UEs supported by the TSP scheme. For 4 UEs, the IMP scheme has a performance gain of more than 3 dB compared with the TSP scheme at BLER=0.1. For 6 to 10 UEs and at SNR less than -2 dB, the performance of the TSP scheme will be slightly better than the IMP scheme due to channel estimation accuracy. With the increasing of SNR, there obviously exists error floor for the TSP scheme. The main reason is the pilot collision probability increases as the number of UEs increases, which seriously affects the performance of the TSP scheme. Thanks to the lower pilot collision probability, the IMP scheme can significantly improve BLER performance and increase number of access UEs.

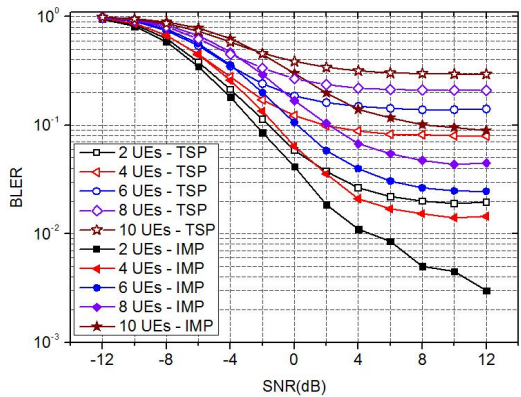

Fig. 5. Comparison of BLER performance between the TSP scheme and the IMP scheme with 2 pilots under 1/7 pilot overhead.

More comparisons of performance between the TSP scheme and the IMP scheme configured with 2 pilots are provided in Fig. 6, where BLER = 0.1 and 0.01 are focused, and different pilot overheads including 1/7 and 2/7 are considered. The performance gain of the IMP scheme shown in Fig. 5 can also be observed in Fig. 6. Due to lower pilot collision probability achieved by increasing pilot overhead, performance of the TSP scheme and the IMP scheme are both improved obviously under 2/7 pilot overhead, and better performance can be provided by the IMP scheme at either BLER = 0.1 or BLER = 0.01.

Fig. 7 shows the performance of the IMP scheme with 2 pilots or 3 pilots under different pilot overhead. The IMP scheme with 3 pilots has similar performance to the case with 2 pilots at BLER = 0.1, while at BLER = 0.01, the IMP scheme with 3 pilots can support more UEs at given SNR or require lower SNR at given number of UEs, especially for the scenario with more access UEs. In other words, the IMP scheme configured with more pilots would have better performance at higher reliability region.

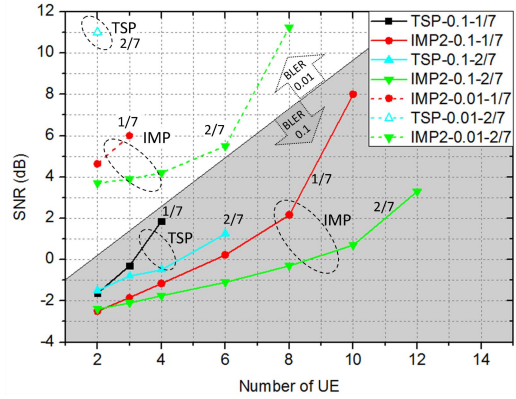

Fig. 6. Performance comparison between the TSP scheme and the IMP scheme with 2 pilots under different pilot overhead.

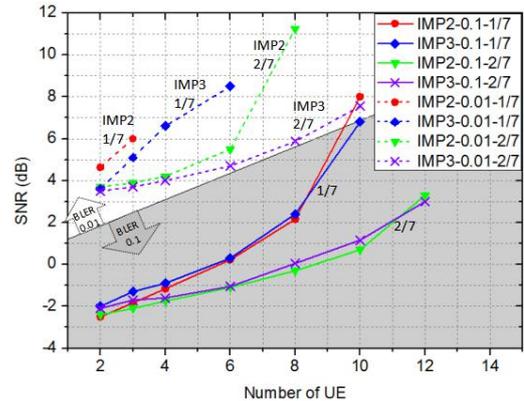

Fig. 7. Performance of the IMP scheme with 2 pilots or 3 pilots under different pilot overhead.

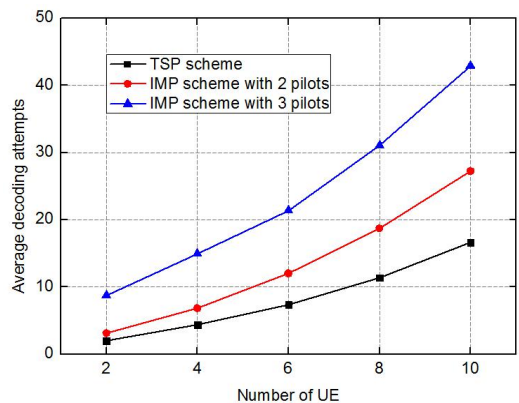

Fig. 8. Comparison of average decoding attempts between the TSP scheme and the IMP scheme.

Fig. 8 shows the comparison of average decoding attempts between the TSP scheme and the IMP scheme with 2 pilots and 3 pilots under 1/7 pilot overhead at SNR=6dB. The decoding attempts of the IMP scheme with 2 pilots is about 1.5 times that of the TSP scheme, and the decoding attempts of the IMP scheme with 3 pilots would increase further. That is to say, the IMP scheme significantly improve the system performance at the cost of increasing the receiver's complexity. Because of lower accuracy of channel estimation, large residual error will be produced and transferred to next round of blind MUD, not only affecting the detection of the remaining UEs, but also leading to some decoded UEs being detected again and the decoding attempts being increased.

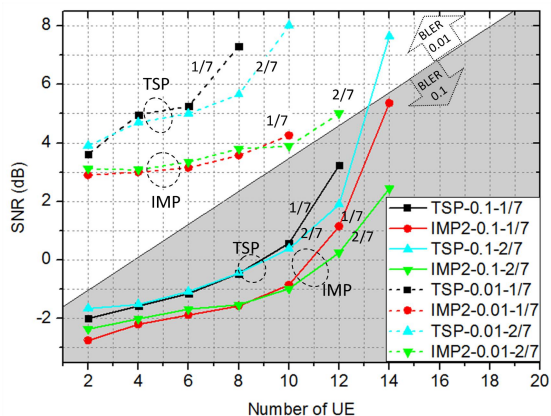

Fig. 9. Performance comparison between the TSP scheme and the IMP scheme with 2 pilots under different pilot overhead.

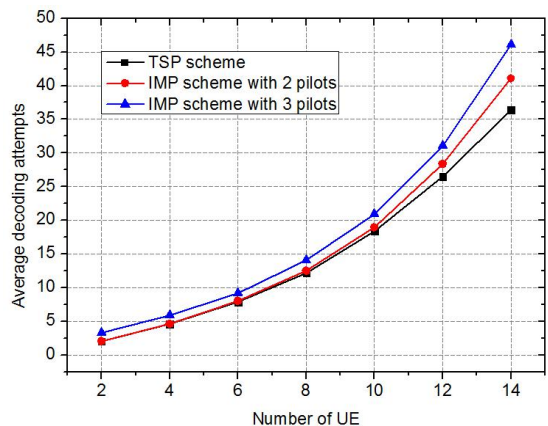

Fig. 10. Comparison of average decoding attempts between the TSP scheme and the IMP scheme.

### B. Data-aided Channel Estimation Used in IC

In this subsection, reconstructed data aided channel estimation is used in IC to reduce the residual error. The performance comparison between the TSP scheme and the IMP scheme configured with 2 pilots under different pilot overhead is shown in Fig. 9. It can be seen that, with data-aided channel estimation, the performances are significantly improved for both the TSP scheme and the IMP scheme compared to the results in Fig. 6. In addition, the IMP scheme with 2 pilots still has a performance gain of at least 2 dB under different pilot overhead at BLER = 0.1 and high UE loading scenario. The performance gain is more obvious at BLER = 0.01. Due to the advantage of data-aided channel estimation, the performance improvement with 2/7 pilot overhead tends to be small, similar conclusion can also be observed for the IMP scheme with 3 pilots, the results of which are not shown in the figure.

The comparison of average decoding attempts between the TSP scheme and the IMP scheme under 1/7 pilot overhead is shown in Fig. 10. With data-aided channel estimation, the average decoding attempts of the IMP scheme is reduced significantly and close to that of the TSP scheme.

## IV. CONCLUSION

A grant-free independent multi-pilot scheme was proposed in this paper, which can efficiently reduce the pilot collision probability. The design of multiple independent pilots and the receiver's flow are introduced in detail. The pilot collision probability, channel estimation accuracy and receiver complexity are analyzed. The simulation results show that independent multi-pilot scheme can significantly improve the BLER performance and increase the number of access UEs. The proposed scheme has potential applications in mMTC scenario in future wireless communications.

REFERENCES

[1] B. Aazhang, P. Ahokangas, H. Alves, M. S. Alouini, J. Beek, H. Benn, M. Bennis, J. Belfiore, E. Strinati, F. Chen, K. Chang, F. Clazzer, S. Dizit, K. DongSeung, M. Giordiani, W. Haselmayr, J. Haapola, E. Hardouin, E. Harjula, and P. Zhu, Key drivers and research challenges for 6G ubiquitous wireless intelligence (white paper), 2019.

[2] H. Shariatmadari, R. Ratasuk, S. Iraji, et al. "Machine-type communications: current status and future perspectives toward 5G systems," IEEE Communications Magazine, vol. 53, no. 9, pp. 10-17, 2015.

[3] H. Xiao, B. Ai, and W. Chen, "A grant-free access and data recovery method for massive machine-type communications," 2019 IEEE International Conference on Communications, 2019.

[4] L. Tian, C. L. Yuan, W. M. Li, et al. "On Uplink Non-Orthogonal Multiple Access for 5G: Opportunities and Challenges", China Communications, vol. 14, no. 12, pp. 142-152, 2017.

[5] D. Tse and P. Viswanath, Fundamentals of Wireless Communication, Cambridge University Press, Cambridge, U.K., 2005.

[6] L. L. Dai, B. C. Wang, Y. F. Yuan, et al. "Non-orthogonal multiple access for 5G: solutions, challenges, opportunities, and future research trend," IEEE Communications Magazine, vol. 59, no. 9, pp. 74-81, 2015.

[7] 3GPP, R1-1609548, "DMRS design for non-orthogonal UL multiple access user channel estimation," ETRI.

[8] H. S. Jang, S. M. Kim, H. S. Park, et al. "An early preamble collision detection scheme based on tagged preambles for cellular M2M random access," IEEE Transactions on Vehicular Technology, vol. 66, no. 7, pp. 5974-5984, 2017.

[9] 3GPP, R1-1611500, "Further consideration on the preamble design for grant-free non-orthogonal MA,", ZTE.

[10] H. S. Jang, S. M. Kim, K. S. Ko, et al. "Spatial Group Based Random Access for M2M Communication," IEEE Communications Letters, vol. 18, no. 6, pp. 961-964, 2014.

[11] A. Quayum, H. Minn, and Y. Kakishima, "Non-orthogonal pilot designs for joint channel estimation and collision detection in grant-free access systems," IEEE Access, vol. 6, pp. 55186-55201, 2018.

[12] Z. F. Yuan, Y. Z. Hu, W. M. Li, et al. "Blind multi-user detection for autonomous grant-free high-overloading multiple-access without reference signal," 2018 IEEE 87th Vehicular Technology Conference, 2018.

[13] Z. F. Yuan, W. M. Li, Y.Z. Hu, et al. "Blind Multi-user Detection based on Receive Beamforming for Autonomous Grant-Free High-Overloading Multiple Access," 2019 IEEE 5G World Forum (5GWF), Dresden, Germany, pp. 526-529, 2019.

[14] H. Jiang, D. Qu, J. Ding and T. Jiang, "Multiple Preambles for High Success Rate of Grant-Free Random Access With Massive MIMO," in IEEE Transactions on Wireless Communications, vol. 18, no. 10, pp. 4779-4789, Oct. 2019